\documentclass[aip,graphicx]{revtex4-1}

\usepackage{color}
\usepackage{graphicx}
\usepackage{amsmath}
\usepackage[utf8]{inputenc}
\usepackage[T1]{fontenc} % Use modern font encodings
\usepackage{dcolumn}% Align table columns on decimal point
\usepackage{bm}% bold math
\usepackage{hyperref}% add hypertext capabilities
\usepackage{array}
\usepackage{textcomp}
\usepackage{multirow}
\usepackage{subscript}
\usepackage{inputenc}
\usepackage{xcolor} %for coloured text
\usepackage[normalem]{ulem}
\usepackage{mathtools}
\usepackage{amsfonts,amsmath,amssymb,bm,empheq}
\usepackage[english]{babel}
\newcommand{\SC}{\textcolor{black}}

\usepackage{soul}  %%\st{text}

\begin{document}

\title{Mn-Intercalated MoSe$_2$ under pressure: electronic structure and vibrational characterization of a dilute magnetic semiconductor}

\author{Shunda Chen}
\email{shdchen@ucdavis.edu}
\affiliation{Department of Chemistry, University of California Davis, One Shields Ave. Davis, CA 95616, USA}
\affiliation{Department of Civil and Environmental Engineering, George Washington University, Washington, DC, 20052, USA}

\author{Virginia L. Johnson}
\affiliation{Department of Chemistry, University of California Davis, One Shields Ave. Davis, CA 95616, USA}

\author{Davide Donadio}
\email{ddonadio@ucdavis.edu}
\affiliation{Department of Chemistry, University of California Davis, One Shields Ave. Davis, CA 95616, USA}

\author{Kristie J. Koski}
\email{koski@ucdavis.edu}
\affiliation{Department of Chemistry, University of California Davis, One Shields Ave. Davis, CA 95616, USA}

\date{\today}

%\keywords{Transition-metal dichalcogenide, MoSe$_2$, Intercalation, Manganese, Diamond anvil cell, Raman scattering, Electronic structure, Pressure effects, Dilute magnetic semiconductor}

 \begin{abstract}
  Intercalation offers a promising way to alter the physical properties of two-dimensional (2D) layered materials. Here we investigate the electronic and vibrational properties of 2D layered MoSe$_2$ intercalated with atomic manganese  at ambient and high pressure up to 7 GPa by Raman scattering and electronic structure calculations. The behavior of optical phonons is studied experimentally with a diamond anvil cell and computationally through density functional theory calculations. Experiment and theory show excellent agreement in optical phonon behavior. The previously Raman inactive A$_{2u}$ mode is activated and enhanced with intercalation and pressure, and a new Raman mode appears upon decompression, indicating a possible onset of a localized structural transition, \SC{involving the  bonding or trapping of intercalant in 2D layered materials}. Density functional theory calculations reveal a shift of Fermi level into the conduction band and spin polarization in Mn$_x$MoSe$_2$ that increases at low Mn concentration and low pressure. Our results suggest that intercalation and pressurization of van der Waals materials may allow one to obtain dilute magnetic  semiconductors with controllable properties, providing a viable route for the development of new materials for spintronic applications.
\end{abstract}

\maketitle


\section{Introduction}
Manganese incorporation in two-dimensional layered and semiconducting materials has received increasing attention as it shows promise for revolutionizing advancements in spintronics \cite{Sato,dietl2010ten,ohno_window_2010,dietl_RMP_2014}, nanostructures with ferromagnetic ordering \cite{Ashwin,mishra2013long,wang2016robust} and tunable functionalities \cite{Robinson,miao2018tunable,wang2016robust}. These perspectives have inspired doping investigations that revealed a plethora of unique magnetic and opto-electronic behaviors. Theory and experiment have shown that Mn doping can lead to ferromagnetic ordering in 2D materials \cite{mishra2013long,miao2018tunable,wang2016robust}. A recent study of the 2D dilute magnetic semiconductor Mn-doped MoS$_2$ predicts that MoS$_2$ doped to 10-15\% manganese is ferromanetic at room temperature \cite{Ashwin}. Intrinsic ferromagnetism in Mn$_x$Mo$_{1-x}$S$_2$ nanosheets, doped by supercritical hydrothermal methods, was reported~\cite{tan2017intrinsic}.  Mn-doping in MoSe$_2$ has also been shown to promote additional active sites for hydrogen evolution reactions~\cite{kuraganti2019manganese}. It was also reported that MnBi$_2$Te$_4$, an intrinsic magnetic topological insulator, is an ideal platform to realize a high-temperature quantum anomalous hall insulator states \cite{lee2018spin}.

In layered materials, Mn intercalation offers a unique alternative to Mn doping. Intercalation, the insertion of an atom or molecule into the van der Waals gap, provides a chemical handle to tune physical properties including electronic structure and phonon propagation without disturbing the host lattice~\cite{Wang,Whittingham,Dresselhaus}. Intercalation in layered materials has demonstrated an enormous realm of physical and chemical tunability in both current and historical research~\cite{Dresselhaus,Schollhorn,Wang,chen,sood_NC_LiMoS2_2018,chen_stronglytunable_2019}. Through intercalation it is possible to adjust the superconducting temperature \cite{Geballe,Dresselhaus}, enhance transparency and conductivity \cite{Wang2,Fuhrer,Cui,Zhu}, and reversibly alter optoelectronic behaviors including color and photoluminescence \cite{Wang3,Fuhrer,Cui}. Recently a wet chemical route was achieved to intercalate zero-valent manganese, post-growth, into 2D layered materials \cite{Wang}, opening a new avenue for experimental study of manganese incorporated two-dimensional materials beyond that of doping.

Molybdenum diselenide (MoSe$_2$) is a heavily investigated layered n-type indirect band gap semiconductor ($E_g$ = 1.1 eV) that shows a transition to a direct band gap with reduction in the number of layers \cite{Kim,Heinz,Strano}. In a recent report, spin states protected from intrinsic electron--phonon coupling were demonstrated in monolayer MoSe$_2$, reaching 100 ns lifetimes at room temperature \cite{ersfeld_spin_2019}. High-pressure investigations have shown that MoSe$_2$ does not undergo any phase transitions up to 30 GPa \cite{Caramazza}. Above 40 GPa, a possible semiconductor-metal phase transition has been identified \cite{Yang,Zhao}. The effect of Mn intercalation on the pressure-induced metallization of MoSe$_2$ is a point of interest, as the intercalated metal may alter the metallization behavior of van der Waals systems \cite{VLjohn}. Whether pressure favors magnetic ordering at ambient temperature is another point of interest as an analogous mechanism was observed with high concentration of substitutional Mn in monolayer MoS$_2$ \cite{Ashwin}. Pressure results in greater wavefunction overlap that could lead to a stronger coupling between isolated Mn intercalant and the MoSe$_2$ semiconductor electron density, enhancing spin polarization effects. Understanding how structure and bonding in this material system change at pressure can provide crucial insight into its fundamental nature.

Here, through both experiment and first-principles calculations, we show that both the intercalation of manganese into MoSe$_2$ and pressurization can alter the host structure and its optical phonon frequencies, giving rise to new Raman-active vibrational modes and modifying the electronic band structure. Pressure-dependent Raman scattering,  investigated up to 7 GPa under hydrostatic conditions in a diamond anvil cell, suggests the formation of pressure-induced bonding between selenium and the manganese intercalant. First-principles calculations exhibit Raman shifts in agreement with experiments and shed light on the changes of the magnetic and vibrational properties of intercalated MoSe$_2$ with low Mn content as a function of pressure. 
In addition, electronic structure calculations provide predictions as for the structural and electronic properties of MoSe$_2$ intercalated with higher amounts of Mn. 
Spin-polarized band structure calculations unravel the conditions at which Mn-intercalated MoSe$_2$ can sustain significant spin currents, making it a suitable dilute magnetic semiconductor.

\section{Methods}   

\subsection{Manganese Intercalation}
Molybdenum diselenide (MoSe$_2$) was prepared by deposition from as-delivered powder containing large single-crystal platelets onto fused silica substrates followed by drop-casting ethanol onto the substrate to adhere the layered material and prevent loss in solution during intercalation. MoSe$_2$ single-crystal platelets were on the order of 1--100 $\mu$m with varied thicknesses ranging from tens of nanometers to microns. 

Zero-valent manganese was intercalated through the decomposition of dimanganese decacarbonyl (C$_{10}$O$_{10}$Mn$_{2}$) in dilute acetone under inert atmosphere \cite{Wang}. This route was shown to successfully intercalate manganese into hosts. For completeness, the MoSe$_2$ coated substrates were placed in a 25-50 mL round bottom flask with a reflux condenser attached to a Schlenk line, evacuated, and flushed with N$_2$ gas. Extra-dry acetone (5 ml) was added to the flask and heated to 48$^{\circ}$C. A 10 mM solution of the carbonyl in 5 ml of acetone was added to the flask dropwise over the course of 1.5 hr and kept at 48$^{\circ}$C for an additional $\sim$ 1 hr. Substrates were then removed from the solution and rinsed with acetone. All chemicals and powders were obtained from Sigma-Aldrich. 

\subsection{High Pressure}
High pressures were generated using an Alamax EasyLab mini-Bragg diamond anvil cell (DAC) with Boehler anvils with 0.6 mm culets. Spring steel gaskets were pre-indented to 80-100 $\mu$m and drilled with a 250 $\mu$m hole. Ruby spheres (Alamax) were used as a pressure calibrant. In the DAC, a solution of 4:1 v/v methanol:ethanol was used as pressure transmitting fluid. Pressures up to 7 GPa were measured to avoid all phase transitions and to remain at relatively hydrostatic pressures of the pressure transmitting fluid. Single crystal platelets were identified optically.

\begin{figure*}[t]
	\begin{center}
		\includegraphics{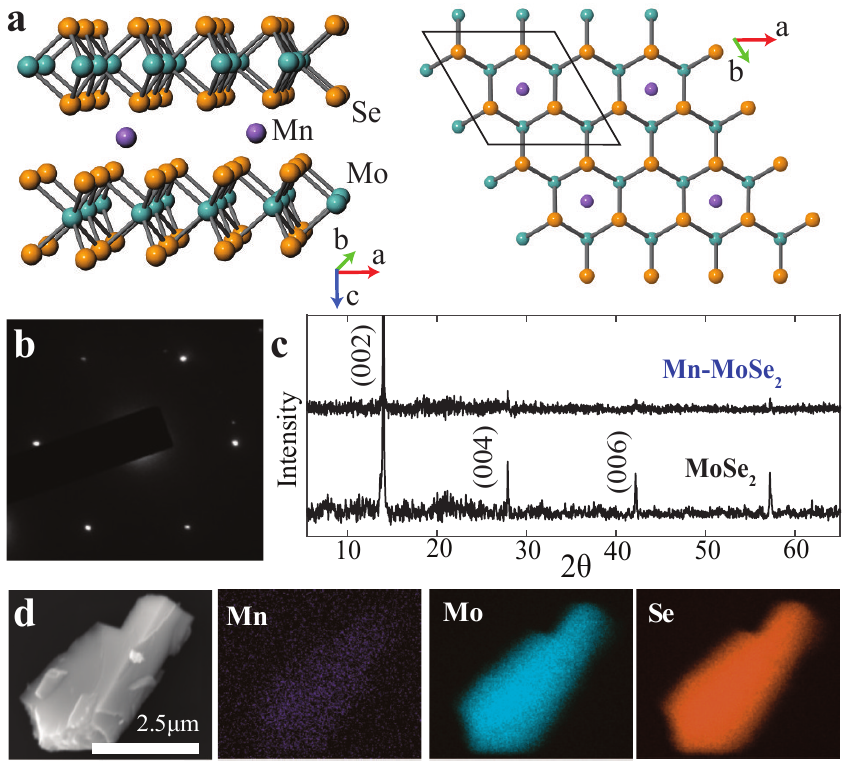}
	\end{center}
	\caption{\label{Fig1} 
		(a) Schematic crystal structure of Mn-MoSe$_2$. Mn likely occupies octahedral sites. 
		(b) SAED of Mn-MoSe$_2$ confirms hexagonal structure with intercalation. 
		(c) XRD show expansion of the MoSe$_2$ host with Mn intercalation. 
		(d) SEM-EDX of Mn-MoSe$_2$ shows Mn throughout.
	}
\end{figure*} 
\subsection{Characterization} 
Raman spectra and ruby fluorescence were measured using a home-built system with a $\lambda$ = 532 nm Coherent Sapphire operating with $<$1 mW on the sample, Leica DMi8 inverted microscope, Princeton Instruments Isoplane SC320, and Princeton Instruments Pixis CCD camera. For Raman spectroscopy, a Semrock laser-edge filter and dichroic with an edge cutoff of $\sim$38 cm$^{-1}$ was used with the same apparatus. Acquisition times were on the order of 5 - 15 seconds with 10 - 20 averaged spectra. All spectra were acquired using an 1800 groove/mm grating. Scanning electron microscopy (SEM) and energy dispersive X-ray spectroscopy (EDX) were acquired on 5-10 isolated flakes with a FEI SCIOS Dual-Beam FIB/SEM using an Oxford EDX detector with 10-20 keV accelerating voltage. Transmission electron microscopy (TEM) and selected area electron diffraction (SAED) were acquired using a JEOL 2100Fac operating at 200 keV. X-ray diffraction (XRD) data was acquired using a Bruker D8 Advance Eco with copper k-edge ($\lambda$ = 1.54 $\mbox{\AA}$) X-rays. Rietveld refinement was performed with GSAS to determine lattice constants \cite{Bindzus}. 

\subsection{First-principles Calculations}
Density functional theory (DFT) calculations were performed within the local spin density approximation (LSDA) of the exchange and correlation functional \cite{perdew_self-interaction_1981} by using the Quantum-Espresso package \cite{QE-2017}.
Core electrons are treated implicitly through projector augmented wave (PAW) pseudopotentials \cite{blochl_projector_1994,kresse_ultrasoft_1999}, and the valence electronic wavefunctions are expanded in a plane-wave basis set with a kinetic energy cutoff of 49 Ry. The charge density is integrated on 16$\times$16$\times$4 Monkhorst-Pack meshes of \emph{k-}points for pristine MoSe$_2$. Structural and cell relaxations are performed using 
a quasi-Newton optimization algorithm with a convergence criterion of $10^{-4}$ Ry/Bohr for maximum residual force component. 
The strong correlation effect of transition metal d-electrons is considered using the LSDA+U method \cite{anisimov_band_1991,cococcioni_linear_2005}, by introducing a Hubbard type interaction. We use a moderate $U_{eff}$ = 4 eV for both Mo and Mn \cite{zhou_ceder_lda+u_2004,wang_ceder_2006,jain_formation_2011,andriotis_mos2_2014,wu_mos2_2018}. Other $U_{eff}$ values, for example, 2 eV and 6 eV, were tested and consistent results were obtained. With LSDA+U, the calculated band gap of pure MoSe$_2$ increases from 0.7 eV to 0.8 eV, with respect to LSDA results. The frequencies of the phonon modes at the Brillouin zone center (Raman shift) as a function of pressure were calculated using density-functional perturbation theory (DFPT) \cite{baroni_phonons_2001}. The threshold for the iterative calculation of the perturbed Kohn-Sham wavefunctions was set to $10^{-16}$ Ry. 
This approach is well suited to predict Raman shift in monolayer and few-layer transition metal dichalcogenides (TMDs) upon strain \cite{rice_raman-scattering_2013,wang_strain-induced_2016} as well as in bulk \cite{wolverson_raman_2014}. 

%To model Mn intercalation of MoSe$_2$ with different Mn concentrations of 12.5\%, 6\% and 3\%, we used the optimized MoSe$_2$ \DD{2H and 1T} structures to construct 2$\times$2$\times$1, 3$\times$3$\times$1, 4$\times$4$\times$1 supercells, and inserted a Mn atom in the vdW gap. For Mn = 25 atomic \%, two Mn atoms were inserted in two separate vdW gaps of a 2$\times$2$\times$1 supercell. We use 4$\times$4$\times$2 \emph{k-}point mesh for 2$\times$2$\times$1 supercell, and 2$\times$2$\times$2 \emph{k-}point mesh for 3$\times$3$\times$1 and 4$\times$4$\times$1 supercells. 

To model Mn intercalation of MoSe$_2$ with different Mn concentrations and to account for possible structural changes induced by intercalation, we considered both 2H and \SC{1T'} MoSe$_2$ initial structures\footnote{The 1T phase for MoSe$_2$ is unstable and intercalated systems relax into a lower-symmetry 1T' phase with a larger unit cell.},
with Mn atom(s) intercalated in the vdW gap (Details are provided in Table S1 in supporting information). \SC{The total and spin-polarized carrier concentrations were calculated from DFT spin-polarized Kohn-Sham states integrating the first Brillouin zone on the uniform k-point meshes at least 27 times denser than those k-point meshes used in the self-consistent calculations, using the BoltzTrap software.\cite{boltztrap2006}}
%
% for the detailed information about the size of supercells for constructing Mn-
%intercalated MoSe2 structures for DFT calculations with various Mn concentrations in 2H and 1T$^\prime$ phases, and the corresponding Monkhorst-Pack k-point Meshes.The 1T phase of Mn-intercalated MoSe$_2$ is unstable, it eventually goes to 1T' phase.)}
%Spin-polarized framework was used to take into account the magnetic properties. 

\section{Results and Discussion}

\subsection{Structure and Raman Scattering} 

\begin{table*}[t!]
%\begin{table*}
	%\begin{center}
		\caption{\label{Table1} Equilibrium lattice parameters of pristine MoSe$_2$ and Mn-intercalated MoSe$_2$ from experiments and DFT calculations. \SC{In parenthesis it is indicated whether Mn is interstitial within the MoSe$_2$ layer or in the vdW gap between two layers.}}
		%\begin{ruledtabular}
		\begin{tabular}{@{}lcccc}
			%\br
			%\thickhline
			\hline
			& $a$ {\AA} & $c$ {\AA } & Vol. {\AA}$^3$ \\
			%\mr
			%\thinhline
			\hline
			MoSe$_2$ (Exp.) & 3.285(3) & 12.921(3) & 120.8(2) \\ %Rwp = 27% 
			2H-MoSe$_2$ (DFT)& 3.260 & 12.718 & 117.0 \\
			\hline
			Mn$_{0.02}$MoSe$_2$ (Exp.)	& 3.336(5) ($+1.5\%$) & 12.940(5) & 124.7(3) \\
			%2H-MnMoSe2
			2H-Mn$_{0.03}$MoSe$_2$ (DFT, \SC{vdW gap})& 3.260 & 12.801 & 117.9\\
			2H-Mn$_{0.06}$MoSe$_2$ (DFT, \SC{vdW gap})& 3.261 & 12.879 & 118.6\\
			2H-Mn$_{0.125}$MoSe$_2$ (DFT, \SC{vdW gap})& 3.261 & 13.065 & 120.3\\
			2H-Mn$_{0.25}$MoSe$_2$	(DFT, \SC{vdW gap})& 3.273 & 13.423 & 124.3 \\
			%1T-Mn$_{1}$MoSe$_2$	(DFT)& 3.332 & 12.757 & 122.7 \\
			1T'-Mn$_{0.06}$MoSe$_2$	(DFT, \SC{vdW gap})& 3.392 & 12.227 & 118.9 \\
			1T'-Mn$_{0.25}$MoSe$_2$	(DFT, \SC{vdW gap})& 3.428 & 12.021 & 121.0 \\
			1T'-Mn$_{1}$MoSe$_2$	(DFT, \SC{vdW gap})& 3.501 & 12.094 & 124.9 \\
			%Interstitial%
\SC{2H-Mn$_{0.03}$MoSe$_2$ (DFT, interstitial) } & \SC{3.283 } & \SC{12.707} & \SC{118.6}\\
\SC{2H-Mn$_{0.06}$MoSe$_2$ (DFT, interstitial)} & \SC{3.299 } & \SC{12.710} & \SC{119.9}\\
\SC{2H-Mn$_{0.125}$MoSe$_2$ (DFT, interstitial) }& \SC{3.353 } & \SC{12.728} & \SC{123.9}\\

% 			%Interstitial%
% \SC{2H-Mn$_{0.03}$MoSe$_2$ (DFT, interstitial) } & \SC{3.283 ($+0.7\%$)} & \SC{12.707} & \SC{118.6}\\
% \SC{2H-Mn$_{0.06}$MoSe$_2$ (DFT, interstitial)} & \SC{3.299 ($+1.2\%$)} & \SC{12.710} & \SC{119.9}\\
% \SC{2H-Mn$_{0.125}$MoSe$_2$ (DFT, interstitial) }& \SC{3.353 ($+2.9\%$)} & \SC{12.728} & \SC{123.9}\\

%\br
%\thickhline
			\hline
		\end{tabular}
		%\end{ruledtabular}
	%\end{center}
\end{table*}

\begin{figure}[htb!]
	\centering
	\includegraphics{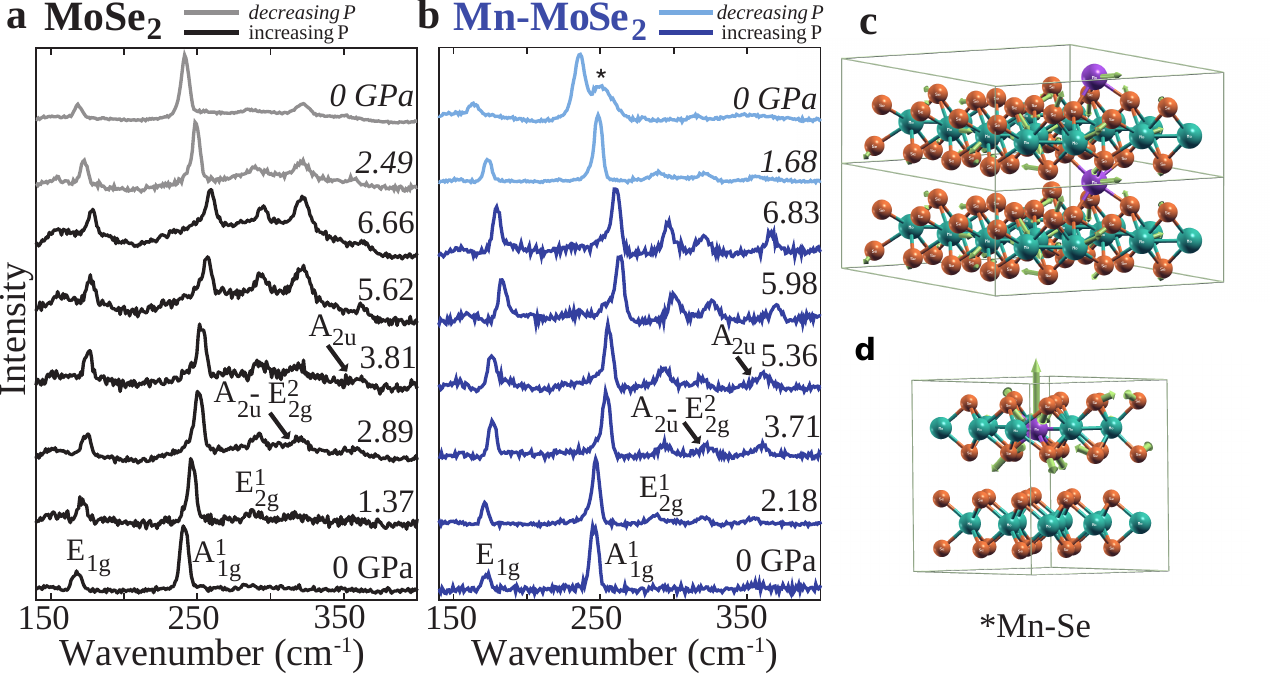}
	\caption{\label{Fig2} (a) Raman spectra under pressure of pristine MoSe$_2$ and (b) Mn-intercalated MoSe$_2$ (black and dark blue lines for increasing pressure, gray and cyan lines for decreasing pressure). The asterisk, described in the text, coincides with the Raman frequency of a Mn-Se bond. (c) Force vectors for the mode at $\sim$ 250 cm$^{-1}$ in 1T$^\prime$-Mn$_{0.06}$MoSe$_2$ from DFT calculations, showing a collective optical mode that involves displacements of Mn, Mo and Se atoms, \SC{possibly} corresponding to the * Mn-Se peak in experiment. \SC{(d) Force vectors for the Mn-Se collective mode at $\sim$ 250 cm$^{-1}$ in 2H-Mn$_{0.06}$MoSe$_2$ with Mn at interstitial site from DFT calculations, which might also correspond to the * Mn-Se peak in experiment.}}
\end{figure}

\begin{figure}[htb!]
	\centering
	\includegraphics{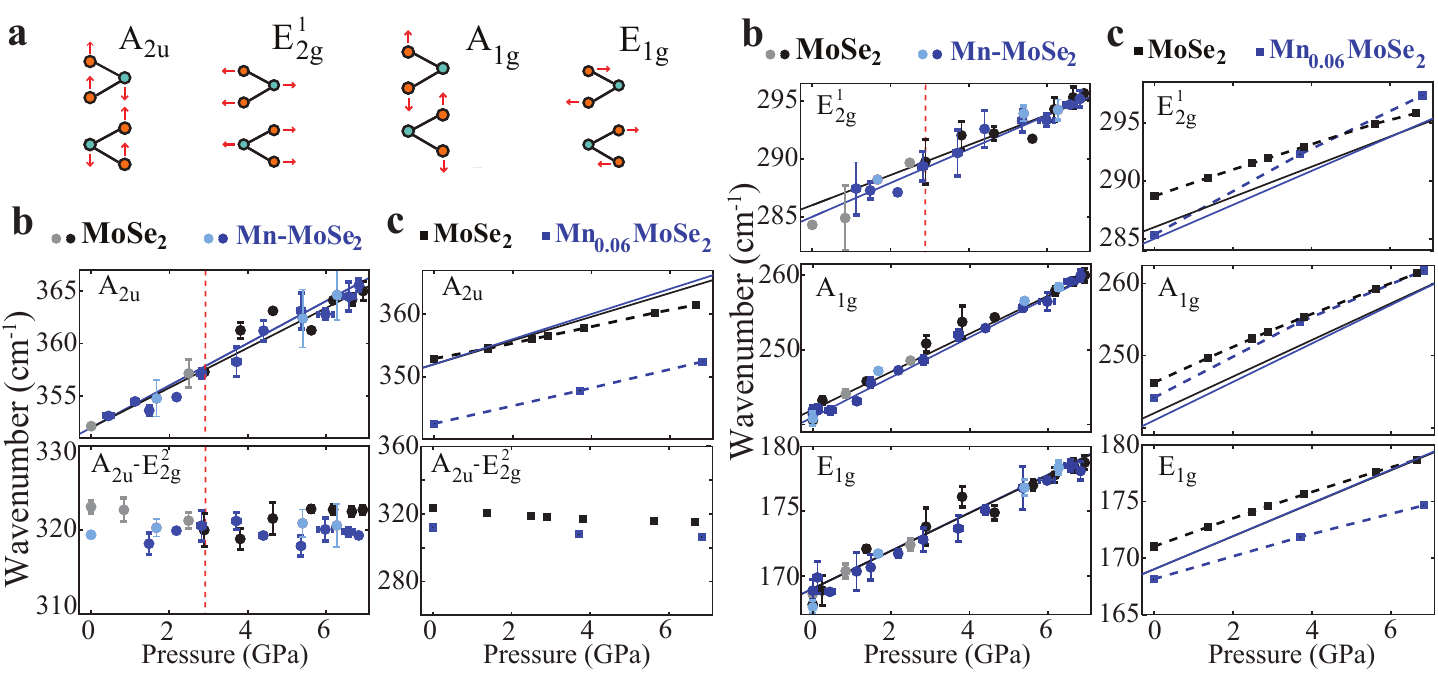}
	\caption{\label{Fig3} (a) Mode schematics. (b) Experimentally determined Raman  shifts for pristine MoSe$_2$ and Mn$_{0.02}$MoSe$_2$ under pressure (black and dark blue full circles for increasing pressure, gray and cyan full circles for decreasing pressure). For pristine MoSe$_2$, A$_{2u}$, A$_{2u}$-E$^{2}_{2g}$, and E$^{1}_{2g}$ modes appear after 2.89 GPa as indicated with a red vertical dashed line. (c) DFT calculated Raman shifts  for pristine MoSe$_2$ and Mn$_{0.06}$MoSe$_2$. The linear fit of the experimental shifts are solid lines for reference.}
\end{figure} 

The 2H stable phase of MoSe$_2$ has a hexagonal crystal structure (Figure \ref{Fig1}(a); space group: \emph{$P6_{3}/mmc$}). The overall host structure is maintained with intercalation as confirmed through SAED (Figure \ref{Fig1}(b)) and XRD (Figure \ref{Fig1}(c)), which show that also Mn-MoSe$_2$ is hexagonal.
An example SEM image with EDX elemental mapping of Mn-intercalated MoSe$_2$ (Figure ~\ref{Fig1}(d)) shows that the intercalant is distributed throughout the plates at concentrations of $\sim$1-2 atomic percent (Mn$_{0.02}$MoSe$_2$). 
Nevertheless, XRD would not be suitable to detect local lattice distortions at the intercalation sites. 

The intercalation energy per Mn atom, estimated by total energy calculations, \SC{depends on Mn concentration,} and it is defined as:
\begin{equation}
	\label{Eq2}
	E_{intercalation} = \frac{E_{Mn_{x}MoSe_2} - E_{MoSe_2} - n_{a}E_{Mn}}{n_{a}}
\end{equation} 
where $E_{Mn_xMoSe_2}$ is the total energy of a MoSe$_2$ supercell intercalated with an atomic concentration $x$ of Mn, $E_{MoSe_2}$ is the total energy of a 2H-MoSe$_2$ supercell with the same number of formula units as the intercalated system, $E_{Mn}$ is the energy per atom of bulk $bcc$ Mn, and $n_a$ is the number of Mn atoms intercalated. 
\SC{The intercalation energy per Mn atom decreases with increasing Mn concentration, from 2.39 eV/atom with at 3\% Mn  to 1.96 eV/atom at 25\% Mn. This trend and the values are similar to copper and silver intercalation in MoS$_2$. \cite{guzman_Cu_Ag_intercalated_MoS2_2017}
Although the intercalation energy per Mn atom decreases slightly with increasing Mn concentration, 
%the number of Mn atoms intercalated per unit cell increases linearly with Mn concentration (please see Table S1 in the supporting information). Therefore, 
the total intercalation energy per MoSe$_2$ unit actually increases as a function of the Mn content from 0.149 eV at 3\% Mn to 0.980 eV per unit cell  at 25\% Mn, indicating that intercalation becomes energetically less favorable the larger the Mn concentration.
\footnote{
The formation energies discussed here are obtained with LSDA+U and are systematically lower than those computed by LSDA, which however exhibit the same trends. Formation energies are summarized in Table S1.}}

In the hypothetical case of $x=1$, we find that the most stable phase of Mn$_1$MoSe$_2$ would be a 1T$^\prime$ with AA stacking (see Figure S2 in supporting information for the phonon dispersion curves--the absence of the imaginary frequency throughout the Brillouin zone indicates the structural stability), as opposed to the AB stacking of the 2H phase. This intercalation-induced structural transition  is analogous to that observed in other metal-intercalated transition metal dichalcogenides, e.g. Li:MoS$_2$ \cite{Eda-2011jt,Cheng-2014fb}. For $x=0.25$ the transition to 1T$^\prime$ lowers the total energy by 0.178 eV per MoSe$_2$ formula unit.
%\SC{\st{, indicating that the 1T$^\prime$ phase is more stable than 2H at high Mn concentration.} 
\SC{Conversely, at lower Mn concentration, for example x=0.125, the 2H phase has lower energy than the 1T$^\prime$ phase by 0.103 eV per MoSe$_2$ formula unit (formation energies per MoSe$_2$ formula unit are summarized in Table S1 in supporting information). 
These calculations predict the 1T$^\prime$ phase to be stable at higher Mn concentration (x$\geq$0.25), and the 2H phase to remain stable at low Mn concentration (x$\leq$0.125). At ambient condition the system would then stay in the 2H phase upon intercalation at the low Mn concentrations of the experiment.}

Successful intercalation of manganese is confirmed by XRD, which shows an expansion of the host lattice constants and the unit cell volume (Table \ref{Table1}; Figure ~\ref{Fig1}(c)). This expansion is associated with insertion of atoms into the van der Waals gap \cite{Wang,Powell}. Expansion of the unit cell volume is measurable, with an almost 3\% increase even at low Mn concentrations. The volume change calculated using DFT is similar with $\sim$ 1-6\% expansion, depending upon the Mn intercalation concentration. 
Experiments with very low Mn concentration show a mild $\sim 0.05$ \AA\ expansion of the in-plane lattice parameter ($a$) upon intercalation and a $\sim 0.02$ \AA\ expansion of the cross-plane lattice parameter ($c$). 
DFT calculations for low concentrations of Mn in the 2H phase would predict an expansion of the $c$-axis only, while $a$ expands at higher concentration of Mn. In contrast, if intercalation is accompanied by a transition to the 1T$^\prime$ structure, $a$ expands and $c$ contracts. While a complete transition to 1T$^\prime$ cannot occur at such low Mn concentration, we argue that the $a$ expansion observed in experiments stems from local distortions around the intercalation sites, which also disrupt the long range crystalline order of the system, as suggested by the disappearance of the high order peaks in XRD (Figure~\ref{Fig1}(c)).
%In contrast, in simulations at slightly higher concentrations only $c$ is affected by Mn intercalation, while $a$ expands at higher concentration of Mn. Such discrepancy between theory and experiments may be ascribed to the inherent approximations used in our DFT calculations. In particular the local density approximation (LDA) used for the exchange and correlation functional is known to underestimate the lattice parameter of covalent systems, and to account for the inter-layer interactions in van der Waals materials only through error cancellation. While we cannot hope to resolve lattice expansions below 0.1 \AA\ by this approach, we show in the following that DFT-LDA provides the correct physical picture of the vibrations of Mn intercalated MoSe$_2$, thus justifying the subsequent discussion about its electronic structure.

Raman spectra of MoSe$_2$ and Mn-MoSe$_2$ as a function of pressure are presented in Figure ~\ref{Fig2}. The observed in-plane modes are E$^2_{1g}$ at 168 cm$^{-1}$ and E$^1_{2g}$ at 286 cm$^{-1}$. The only Raman active out-of-plane mode is the A$_{1g}$ mode, which is initially at 242 cm$^{-1}$ \cite{Caramazza}. A peak at 354 cm$^{-1}$ is seen at pressures above 2.89 GPa and can be assigned as A$_{2u}$ \cite{Kim}. It is an infrared active phonon that is forbidden in Raman scattering \cite{agnihotri1973laser}, as confirmed by DFT calculations. Previous studies observe this peak at higher excitation energies at ambient conditions and reason that resonance effects allow this peak to be observable \cite{Cheong,Kim}. This mode has also been observed at higher pressures \cite{Yang,Zhao}. Intercalation of manganese and application of pressure may result in symmetry breaking allowing the forbidden mode to appear in Raman spectra. Symmetry breaking may also be responsible for the appearance of the E$^1_{2g}$ mode at higher pressures. With increasing pressure, both modes increase in scattering intensity \cite{Caramazza,Yang,Zhao}. Symmetry breaking with pressure is common and not unexpected. These additional peaks may have been observed in these high pressure studies over other investigations because of the high resolution grating used.

An additional feature not observed in MoSe$_2$ appears post decompression at approximately 250 cm$^{-1}$ in Mn-MoSe$_2$ (labeled * in Figure ~\ref{Fig2}(a)). The frequency of this mode matches a calculated longitudinal optical phonon mode in wurtzite MnSe, suggesting bonding between the Mn intercalant and the host with released pressure.\cite{Shen} 
\SC{Through DFT calculations we found that this mode is related to the formation of a bond between the host structure and the intercalant guest. This may, however, happen in different ways: either through a transition to the metastable 1T' structure or through the interstitial embedding of Mn into a MoSe$_2$ layer, which was recently proposed in MoSe$_2$ monolayer on the basis of DFT calculations.\cite{onofrio_novel_2017}}

In fact, the 250 cm$^{-1}$ peak appears in the calculation of 1T$^\prime$ Mn-intercalated MoSe$_2$ at any given concentration of intercalant from 0.06 to 0.25. The frequency corresponds to a collective optical mode that involves displacements of both Mn, Mo and Se atoms (Figure \ref{Fig2}(c) and Figures S3-S4). 
\SC{Since Mn atoms into interstitial sites might be another possible explanation for the onset of the new mode observed in experiment, we computed the phonon frequencies of Mn-MoSe$_2$ with Mn at the interstitial site at 12.5\%, 6\% and 3\%  Mn concentrations. %using 2x2x1, 3x3x1 and 4x4x1 supercells, respectively. 
With Mn at the interstitial site, the calculated in-plane lattice expansion rates are 2.9\%, 1.2\% and 0.7\% for Mn concentrations at 12.5\%, 6\% and 3\%, respectively (see Table \ref{Table1} for the optimized lattice parameters). The calculated in-plane expansion rates for Mn at interstitial are in good agreement with the in-plane lattice expansion rate 1.5\% measured in experiment at low Mn concentration of ~1-2\%. However, the frequency of the Mn-Se collective mode is sensitive to the Mn concentration. At 12.5\% of Mn, the calculated Mn-Se collective mode appears at $\sim$240 cm$^{-1}$ (see Figure S7), which is lower than new mode observed at 250 cm$^{-1}$ in the experiment. And at 6\% Mn concentration, the calculated Mn-Se collective mode appears at around 250 cm$^{-1}$, in very good agreement with the measured value (see Figure \ref{Fig2}(d) and Figure S8). But at 3\% Mn concentration, the Mn-Se collective vibrational mode blue-shifts to around 256 cm$^{-1}$ (see Figure S9 in supporting information). Given the linear shift of the frequency of Mn-Se mode with Mn concentration, one could not conclude that the new mode at 250 cm$^{-1}$ observed in experiment is solely due to the interstitial trapping of Mn. 
Since the 250 cm$^{-1}$ peak appears in the calculation of 1T$^\prime$ Mn-intercalated MoSe$_2$ at any given concentration, we would conjecture that both (i) Mn bound in the vdW gap and (ii) Mn bound in interstitials are possible. However, it remains open whether Mn binds more in the vdW gap or in interstitials and there is no simple experimental route to directly address this hypothesis. These results suggest that Mn intercalation combined with compression and decompression processes may provide possible new routes to  Mn interstitial doping of layered materials. \cite{onofrio_novel_2017,karthikeyan_which_2019}}

\begin{center}
\begin{table*}
\caption{\label{Table2} Ambient pressure Raman frequencies ($cm^{-1}$), Raman shift pressure-derivative, $\frac{d\omega}{dP}$ ($\frac{cm^{-1}}{GPa}$), and isothermal mode Gr\"{u}neisen ($\gamma_T$) for observed modes calculated from Equation (\ref{gruneisen}).}
%\begin{indented}
%\tiny
%\item[]
\begin{ruledtabular}
\begin{tabular}{@{}lcccccc}
%\br
 & \multicolumn{3}{c}{{\bf E$_{1g}$}} & \multicolumn{3}{c}{{\bf A$^1_{1g}$}} \\
 & $\omega_o$ (cm$^{-1}$) & $\frac{d\omega}{dP}$ $\left(\frac{cm^{-1}}{GPa}\right)$ & $\gamma_T$ & $\omega_o$ (cm$^{-1}$) & $\frac{d\omega}{dP}$ $\left(\frac{cm^{-1}}{GPa}\right)$ & $\gamma_T$ \\
\hline
MoSe$_2$ ({Exp.}) & 167.8(6) & 1.68(8) & 0.46(2) & 240.6(6) & 2.9(1) & 0.54(2)\\
\SC{MoSe$_2$ ({DFT})} & 171.02 & 1.15(2) & 0.307(5) & 246.24 & 2.24(4) & 0.416(7)\\
\hline
Mn$_{0.02}$MoSe$_2$ ({Exp.}) & 167.6(6) & 1.67(6) & 0.46(2) & 241.3(6) & 2.67(4) & 0.506(8)\\
\SC{Mn$_{0.06}$MoSe$_2$ ({DFT})} & 168.20 & 0.96(7) & 0.261(2) & 244.44 & 2.6(4) & 0.486(3) \\
\SC{Mn$_{0.125}$MoSe$_2$ ({DFT})} & 165.52 & 0.90(3) & 0.248(8) & 241.45 & 1.88(5) & 0.55(1) \\
\SC{Mn$_{0.25}$MoSe$_2$ ({DFT})} & 158.16 & 0.5(2) & 0.144(1) & 207.21 & 2.1(2) & 0.463(3) \\

\hline

 &\multicolumn{3}{c}{{\bf E$^1_{2g}$}} & \multicolumn{3}{c}{{\bf A$_{2u}$}} \\
 & $\omega_o$ (cm$^{-1}$) & $\frac{d\omega}{dP}$ $\left(\frac{cm^{-1}}{GPa}\right)$ & $\gamma_{T}$ & $\omega_o$ (cm$^{-1}$) & $\frac{d\omega}{dP}$ $\left(\frac{cm^{-1}}{GPa}\right)$ & $\gamma_T$ \\ 
\hline
MoSe$_2$ ({Exp.}) & 286(1) & 1.3(2) & 0.21(3) & 354.2(2) & 1.5(4)  & 0.20(5)\\
\SC{MoSe$_2$ ({DFT})} & 289 & 1.08 & 0.171(1) & 352.8 & 1.31 & 0.215(1)\\
\hline
Mn$_{0.02}$MoSe$_2$  ({Exp.}) & 285.2(5) & 1.5(1) & 0.23(2) & 351.5(5) & 2.0(1) & 0.26(1)\\
\SC{Mn$_{0.06}$MoSe$_2$ ({DFT})} & 285.48 & 1.8(1) & 0.288(2) & 342.45 & 1.45(3) & 0.194(1) \\
\SC{Mn$_{0.125}$MoSe$_2$ ({DFT})} & 279 & 1.29 & 0.240(2) & 331 & 0.899 & 0.1200(8) \\
\SC{Mn$_{0.25}$MoSe$_2$ ({DFT})} & 269.76 & 1.4(2) & 0.237(2) & 316.01 & 0.1(4) & 0.0145(1) \\
\end{tabular}
\end{ruledtabular}
%\end{indented}
\end{table*}
\end{center}

Figure ~\ref{Fig3} shows the measured (Figure ~\ref{Fig3}(b)) and calculated (Figure ~\ref{Fig3}(c)) vibrational frequencies for each Raman mode as a function of pressure. Figure ~\ref{Fig3}(a) shows the schematic of each vibrational mode. In Figure ~\ref{Fig3}(c), calculated Raman shifts are plotted alongside the linear fit of the experimental shifts in solid black and blue line for MoSe$_2$ and Mn-MoSe$_2$, respectively. DFT calculated Raman shifts are in very good agreement with experimental results. Though experimental data do not show significant Raman shifts upon Mn intercalation with low concentrations of Mn intercalant of $\sim$1-2 atomic \%, DFT calculations suggest that the frequency of the Raman shift would decrease with higher Mn concentration at ambient and relatively low pressure (please see Figure ~\ref{Fig3},  and Figure S1 in supporting information).  

Table~\ref{Table2} provides the initial Raman frequency, $\omega_o$, at ambient pressure and the change in frequency with pressure ($\frac{d\omega}{dP}$). Despite a detectable change in the host unit cell volume (Table \ref{Table1}), with low concentrations of Mn intercalant of $\sim$1-2 atomic \%, the experimentally measured Raman frequency shift of each mode in MoSe$_2$ does not change significantly upon Mn intercalation (Table \ref{Table2}). This is not unusual. Experimentally, shifts of the Raman modes with intercalation are complex \cite{reed2019chemically}. Optical phonons can exhibit stiffening, softening, or no change with intercalation affected by the acceptor or donor nature of the intercalant as well as the associated change in the host volume with intercalation \cite{reed2019chemically}.  

From the experimental data, the frequency of the Raman  shift of each mode tends to decrease slightly, except the A$_{1g}$ mode. Though the frequency of the Raman shift of the A$_{1g}$ mode seems to increase by about 0.7 $cm^{-1}$ (Table \ref{Table2}), the measurement error is comparable to the size of the change. From DFT calculations, with higher Mn concentration, it looks more likely that the frequency of the Raman shift of each mode would decrease upon Mn intercalation at ambient and relatively low pressure (see Figure ~\ref{Fig3} and Figure S1 in supporting information). The experimental Raman shift of Mn-MoSe$_2$ and MoSe$_2$ modes show very similar linear pressure-dependent slope (\(\frac{d\omega}{dP}\)) (Figure ~\ref{Fig3}, Table \ref{Table2}). For pristine MoSe$_2$, the E$^1_{2g}$ and A$_{2u}$ modes do not show up until pressure of $\sim$ 2.89 GPa and remain with decompression \cite{agnihotri1973laser,sekine1980raman}. An anomalous mode shows up around 320 cm$^{-1}$. DFT calculations suggest this might be ascribed to the A$_{2u}$-E$^2_{2g}$ combination band. This mode does not appear in MoSe$_2$ until about $\sim$ 2.89 GPa, as indicated by a dashed red line in Figure \ref{Fig3}. This mode persists with decreasing pressure, which is consistent with the appearance of the A$_{2u}$ mode. It occurs in Mn-MoSe$_2$ at around $\sim$ 1 GPa. This peak shows no change with pressure as Raman shifts of both the E$^2_{2g}$ and the A$_{2u}$ increase. Thus, there is no significant pressure-derivative of this mode. All modes show phonon stiffening, increasing linearly with pressure, except the overtone mode A$_{2u}$-E$^2_{2g}$ at around 320 cm$^{-1}$ as discussed above.

Compressibility of MoSe$_2$ and MnMoSe$_2$ can be described using the isothermal mode Gr\"{u}neisen parameter ($\gamma_T$):
\begin{equation} \label{gruneisen}
	\gamma_T = -\left( \frac{d \ln \omega}{d \ln V}\right)_T = \frac{B_T}{\omega_0}\left(\frac{d \omega}{dP}\right)_T
\end{equation}
where $B_T$ is the isothermal bulk modulus. Using the third-order Birch-Murnaghan equation of state to fit \emph{in situ} high-pressure MoSe$_2$ X-ray diffraction data, Aksoy\cite{Aksoy} \textit{et al.} calculated $B_T$ as 45.7 $\pm$ 0.3 GPa. DFT calculations here find a bulk modulus of 47.9 GPa for MoSe$_2$, close to experiments \cite{Aksoy}, and 51.3 GPa for Mn$_{0.125}$MoSe$_2$. Mn-intercalation should yield a notable decrease in the isothermal compressibility. With pressure, the empty van der Waals gap should compress first. By adding more atoms to the gap, Mn-intercalation subsumes space otherwise available for compression. Thus, intercalation makes the material less compressible. 
Using these values of $B_T$, along with the relevant values of $\omega_0$ and $\frac{d\omega}{dP}$, the mode Gr\"{u}neisen is calculated from Equation \ref{gruneisen} for all modes (Table ~\ref{Table2}).  
Calculated isothermal mode Gr\"{u}neisen parameters in most cases exhibit similar trends as experiments, however with large uncertainties, mostly due to shortcoming related to the approximated density functional.  
%turn out in excellent agreement with the measurements. 
However, the overall agreement between Raman measurements and DFT calculations suggests that the adopted level of theory accounts well for the charge redistribution upon intercalation, and DFT calculations can be used to predict the electronic structure of Mn-intercalated MoSe$_2$.

\subsection{Electronic Band Structure}

\begin{figure*}[tb!]
	\centering
	\includegraphics{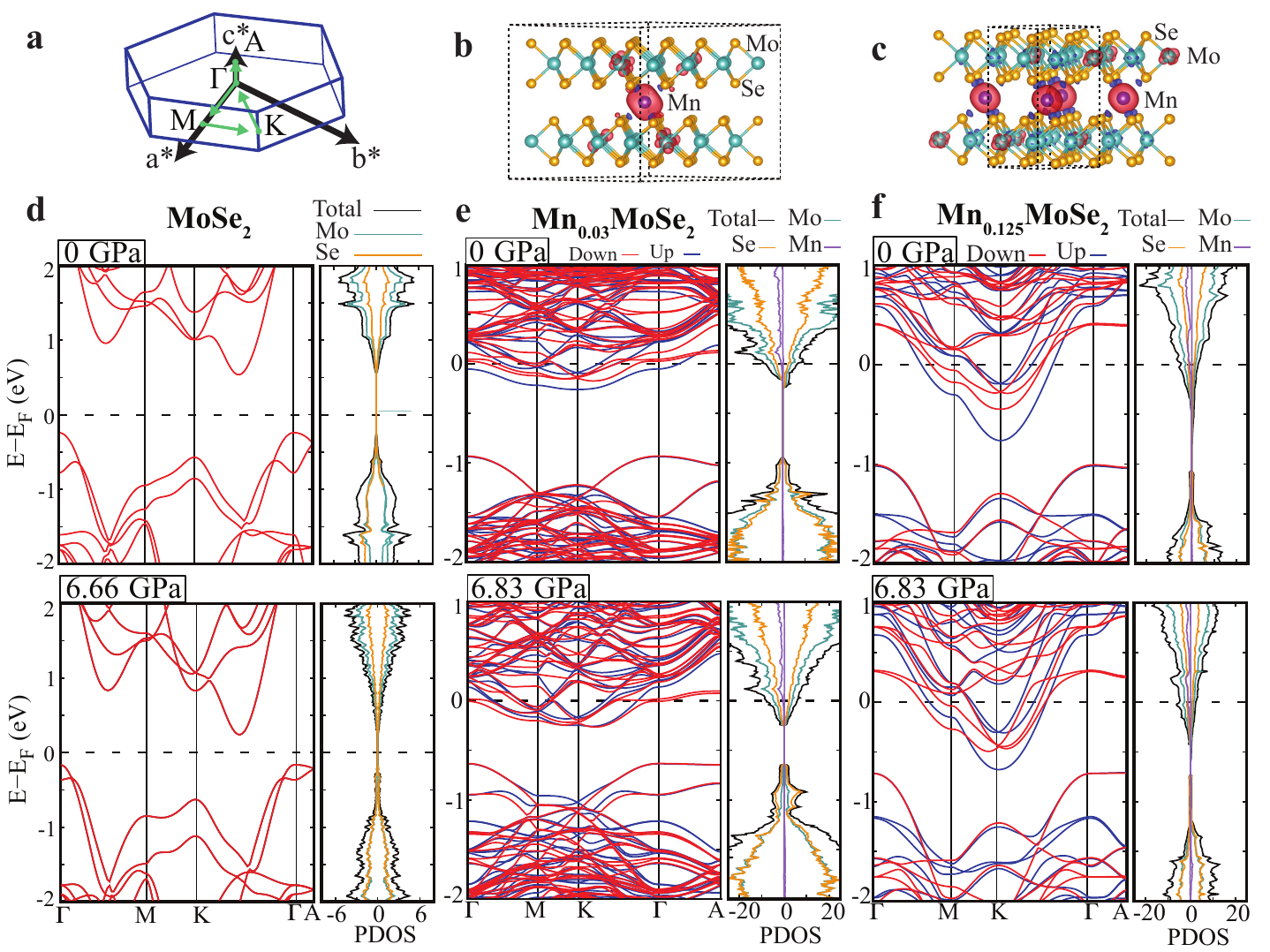}
	\caption{\label{Fig4} (a) Brillouin zone schematic of MoSe$_2$. (b) Mn$_{0.03}$MoSe$_2$ structure and (c) Mn$_{0.125}$MoSe$_2$ structure at 0 GPa with spin density and polarization; red and blue isosurfaces represent positive and negative spin density, respectively. (d) Electronic band structure of MoSe$_2$ at 0 GPa (above) and 6.66 GPa (below). Projected density of states are given for either pressure to the right of the band structure. The band structure and PDOS of (e) Mn$_{0.03}$MoSe$_2$ and (f) Mn$_{0.125}$MoSe$_2$ show that Mn intercalation metallizes the material moving the Fermi level into the conduction band. Manganese states are spin polarized in the conduction band and the population of states increases with pressure as the band gap narrows.}
\end{figure*}

\begin{figure}[htb!]
	\centering
	\includegraphics{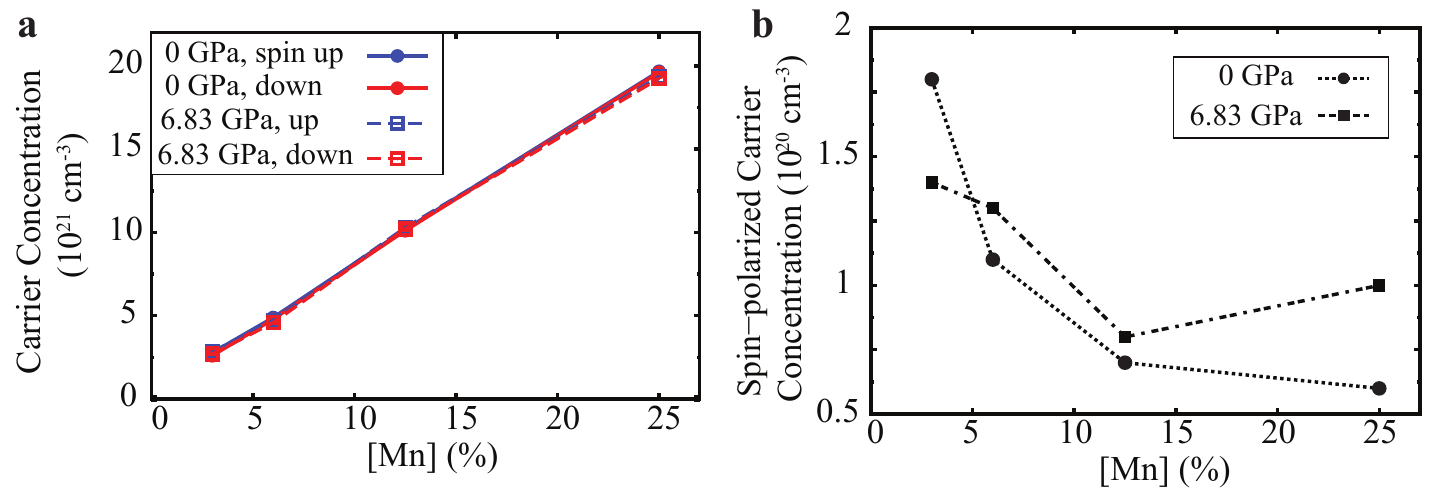}
	\caption{\label{Fig5} (a) Spin up and spin down carrier concentration as a function of Mn concentration at 0 GPa and 6.83 GPa. (b) Total spin-polarized carrier concentration (the difference between spin up and spin down carrier concentrations).}
\end{figure}

Figure~\ref{Fig4} shows the electronic band structures of pristine MoSe$_2$, Mn$_{0.03}$MoSe$_2$ and Mn$_{0.125}$MoSe$_2$ at 0 GPa and $\sim$ 7 GPa. For pristine MoSe$_2$, with the increase of pressure, the band gap narrows from 0.8 eV at 0 GPa to 0.4 eV at 6.66 GPa (Figure \ref{Fig4}(d)). Upon intercalation of Mn, the overall host structure is retained (Figure \ref{Fig4}(b) and (c)). 
At higher concentrations of Mn intercalant (Mn$_{0.125}$MoSe$_2$, Figure \ref{Fig4}(c)) the Fermi level lies deeper in the conduction band. While pressure tends to close the gap between the valence and the conduction band also in intercalated systems, the position of the Fermi level and the carriers concentration are determined by doping and do not change significantly upon compression (Figure \ref{Fig5}(a)).
The transition to the 1T$^\prime$ phase would lead to the metallization of the system, with substantial change of the band structures and spin-polarized density of states. (See Figure S5 in supporting information for the spin-polarized electronic band structure for 1T$^\prime$ phase Mn-intercalated MoSe$_2$.) 
\SC{Conversely, when Mn gets trapped in the interstitial of the MoSe$_2$ layer, the system remains semiconducting with a localized spin state in the gap, below the Fermi level (Figure S10).}% in supporting information for the spin-polarized electronic band structure and spin-polarized projected density of states).}

The calculated projected density of states (PDOS) are shown in Figure \ref{Fig4} for pristine and Mn-intercalated MoSe$_2$ at 0 GPa and at $\sim$ 7 GPa. Mn intercalation in MoSe$_2$ shows clear signatures of spin polarization. At low concentrations of Mn (Mn$_{0.03}$MoSe$_2$), there is clear separation of the density of spin-up and spin-down electrons which suggest spin polarized current is possible in Mn-intercalated MoSe$_2$. With pressure, the spin separation between the spin-up and spin-down states is reduced. 
At 0 GPa the overall magnetic moment of the Mn$_{0.03}$MoSe$_2$ and Mn$_{0.125}$MoSe$_2$ supercell is 5.00 $\mu_B$. With pressure (at 6.83 GPa), the overall magnetic moments reduce slightly to 4.61 $\mu_B$ and 4.54 $\mu_B$ for Mn$_{0.03}$MoSe$_2$ and Mn$_{0.125}$MoSe$_2$, respectively.

The calculated total (n-type) carrier concentrations increase with increasing Mn, proportional to Mn concentration, and do not shift significantly with pressure (Figure \ref{Fig5}(a)). 2D semiconductors with dilute magnetic manganese impurities should have conduction completely ruled by spin polarization \cite{Ashwin,Sato}. One would expect that with increasing manganese concentration and increasing pressure, the concentration of spin-polarized carriers would also change with competing effects due to spin overlap. The spin interaction should increase as the manganese concentration increases (Figure \ref{Fig4}(c)), thus decreasing the spin-polarized carrier concentration. As pressure is increased, one would expect a similar effect, as the spin overlap and spin interaction also increase. Calculations of the spin-polarized carrier concentration as a function of pressure and manganese concentration (Figure \ref{Fig5}(b)) reveal that the concentration of net spin-polarized carriers depends on both pressure and Mn concentration. Small concentrations of Mn show the greatest amount of spin-polarized carriers, with 3 atomic \% (close to that achieved by experiment) as a maximum. It is interesting that at low concentration of Mn (3\%), pressure reduces the spin-polarized carrier concentration, while at higher concentration (above approximately 6\%), pressure increases the spin-polarized carrier concentration (Figure \ref{Fig5}(b)). 
The local structural transition, suggested by the appearance of the 250 cm$^{-1}$ peak in the Raman spectrum upon decompression, would substantially affect the electronic structure of the intercalated material and spin separation.
(See Figure S5 for the band structure and spin-polarized density of states of 1T$^\prime$ phase of Mn-intercalated MoSe$_2$, and Figure S6 for the spin-polarized carrier concentration. 
It’s  interesting that both 2H and 1T$^\prime$ phases show similar trends in spin polarization as a function of Mn concentration, and the spin-polarized carrier concentration for 1T$^\prime$ phase of Mn-intercalated MoSe$_2$ could reach up to $\sim 1\cdot10^{20}$ cm$^{-3}$ at around 11\% of Mn concentration.)
These competing effects reveal the chemical and thermodynamic tunability of MoSe$_2$ spin-polarized carriers. The predicted concentration of spin polarized carriers, up to $\sim 2\cdot10^{20}$ cm$^{-3}$ in 2H phase and $\sim 1\cdot10^{20}$ cm$^{-3}$ in 1T$^\prime$ phase, could possibly be observed by Hall measurements and is significantly high to enable spintronic applications.

\section{Conclusions}
This work illustrates the ability to adjust the phonon frequencies and the electronic band structure with Mn intercalation and pressure. The appearance of a new phase is found associated with Mn guest bonding with the host MoSe$_2$. These results suggest intercalation systems under high pressure can lead to unique bonding environments and, thus, new materials. These findings set precedent for further investigation into Mn-intercalation of 2D layered materials as an alternative to dilute magnetic doping. The robustness of this system is demonstrated by both intercalated and non-intercalated pressure-dependent phonon frequencies.  DFT calculations show that Mn-intercalation causes the Fermi level to shift into the conduction band, rendering the system an n-type semiconductor or nearly metallic. Manganese intercalated MoSe$_2$ retains a total magnetic moment that corresponds to that of single Mn atoms. Unpaired spins contribute to the density of states near the Fermi level, thus potentially enabling spin currents. Pressure reduces the spin carrier density at low Mn concentration, but it slightly increases it at higher Mn concentration. The spin polarized behavior predicted in intercalated Mn-MoSe$_2$ here has the potential to surpass that of doped systems, with the advantage that transition metal atoms may be intercalated post-growth. These results provide insights into how concentration limitations in dilute manganese doped MoSe$_2$ may be bypassed by exploiting the van der Waals gap of a layered material through intercalation and high pressure. We expect that resistivity studies as a function of pressure and magnetic field may further elucidate the nature of Mn spin-polarized carriers in this host.

\section*{Supplementary Material}
{DFT calculated Raman shifts with higher Mn concentrations shown as a function of pressure and Mn concentration, in comparison with experimentally determined Raman shifts; detailed information for constructing Mn-intercalated MoSe$_2$ structures for DFT calculations with various Mn concentrations in 2H and 1T$^\prime$ phases \SC{with Mn in vdW gap, as well as 2H phase with Mn at interstitial site of MoSe$_2$}, and the corresponding Monkhorst-Pack $k$-point meshes \SC{and formation energy}; phonon dispersion  curves  for 1T'-Mn$_1$MoSe$_2$; force vectors of the mode at $\sim$ 250 cm$^{-1}$ in 1T$^\prime$-Mn$_{0.06}$MoSe$_2$ and 1T$^\prime$-Mn$_{0.25}$MoSe$_2$; spin-polarized electronic band structure and PDOS of the 1T$^\prime$ phase of Mn-intercalated MoSe$_2$ at 0 GPa; total spin-polarized carrier concentration for 1T' phase of Mn-intercalated MoSe$_2$ as a function of Mn concentration at 0 GPa and 6.83 GPa; \SC{force vectors of the Mn-Se collective mode in 2H-Mn$_{0.125}$MoSe$_2$, 2H-Mn$_{0.06}$MoSe$_2$ and 2H-Mn$_{0.03}$MoSe$_2$ with Mn at interstial site of MoSe$_2$; spin-polarized electronic band structure and PDOS of 2H-Mn$_{0.03}$MoSe$_2$ with Mn at interstial site of MoSe$_2$.}}

%\begin{suppinfo}

%\end{suppinfo}

\SC{\section*{Data availability}
The data that support the findings of this study are available from the corresponding author upon reasonable request.
}
%%%%%%%%%%%%%%%%%%%%%%%%%%%%%%%%%%%%%%%%%%%%%%%%%%%%%%%%%%%%%%%%%%%%%
\begin{acknowledgments}
We thank Bryan P. Moser and Daniel R. Williams for XRD patterns and SEM images, respectively. This work was supported by the Office of Naval Research (N00014-16-1-3161). 
\end{acknowledgments}

\bibliography{References}

\end{document}